\documentclass[12pt,a4paper,final]{revtex4-1}
\usepackage{amsmath,amssymb}
\usepackage[utf8]{inputenc}
\usepackage{graphicx}
\usepackage{color}
\begin{document}
\title{Ion energy distribution functions behind the
sheaths of magnetized and non magnetized radio
frequency discharges}
\author{Jan Trieschmann, Mohammed Shihab, Daniel Szeremley, 
Abd Elfattah Elgendy, Sara Gallian, Denis Eremin,
Ralf Peter Brinkmann, and Thomas Mussenbrock}
\affiliation{Ruhr University Bochum, Department of 
Electrical Engineering and Information Sciences, 
Institute of Theoretical Electrical Engineering, 
D-44780 Bochum, Germany}
\date{\today}

\begin{abstract}
The effect of a magnetic field on the characteristics of capacitively coupled radio frequency discharges is investigated 
and found to be substantial. A one-dimensional particle-in-cell simulation shows that geometrically
symmetric discharges can be asymmetrized by applying a spatially inhomogeneous
magnetic field. This effect is similar to the recently discovered electrical
asymmetry effect. Both  effects act independently, they can work in the same direction or compensate each other. 
Also the ion energy distribution functions at the electrodes are strongly affected by the magnetic field, 
although only indirectly. The field influences not the dynamics of the sheath itself but rather its operating conditions,
i.e., the ion flux through it and voltage drop across it. To support this interpretation, the particle-in-cell results 
are compared with the outcome of the recently proposed ensemble-in-spacetime algorithm. Although that scheme resolves only the sheath  
and neglects magnetization, it is able to reproduce the ion energy distribution functions with very good accuracy,
regardless of whether the discharge is magnetized or not.

\end{abstract}

\maketitle

\newpage

\section{Introduction}\label{sec:introduction}

In the past three decades plasma processing such as
surface activation, deposition, or etching has become
of great importance for a large number of technological 
applications. With the goal of optimizing plasma 
processes, several types of plasma sources have been 
successfully introduced and studied, e.g., hybrid plasma
sources, capacitively coupled radio frequency discharges
(CCRFDs) driven by more than one frequencies, or magnetized
CCRFDs which are used for magnetically enhanced reactive ion 
etching (MERIE). It has been recently shown that dual
frequency CCRFDs which exploit the electrical asymmetry
effect (EAE) are most suitable for a separate
control of the ion flux and energy at substrates or electrodes
\cite{heil2008, czarnetzki2009}. It is possible to control the
ion energy distribution function over a wide range of parameters.
In fact, this is one of the main goals in the context of plasma
processing \cite{kawamura1999, mussenbrock2012}. 
By utilizing the EAE it is even possible
to invert the electrode functionality, i.e., to change the
DC self bias voltage from a large negative value to a large
positive value even for geometrically symmetric CCRFDs
with the driven electrode area equal to the grounded area.
It has been further shown that geometrically asymmetric
CCRFDs (small driven electrode and large grounded area, or
vice versa), which produce a natural DC self bias voltage,
can be completely symmetrized, in the sense that the natural 
DC self bias voltage can be completely compensated
\cite{schulze2011, schuengel2012}.

In this work we study the effect of an externally
imposed static magnetic field in combination with the
EAE on the characteristics of dual frequency CCRFDs. We 
focus on the sheath dynamics and the IEDFs at the electrodes.
Similar to the MERIE concept where usually a static and spatially 
inhomogeneous magnetic field is applied \cite{cheng1989,
lieberman1991, graves1994, rauf2003, kushner2003, oster2004}, we
investigate the effect of a spatially inhomogeneous 
magnetic field on the IEDF. It will be shown, that the
applied magnetic field is able to (over-)compensate or even to 
amplify the EAE.

The manuscript is organized as follows. After a short 
introduction of the discharge configuration and the
operation regime, a brief review of the employed numerical
schemes, namely a one-dimensional Particle-in-Cell algorithm
using Monte-Carlo collisions (PIC) and the Ensemble-in-Space time
scheme (EST), is presented \cite{shihab2012}. We discuss
simulation results for both the magnetized and the non-magnetized case,
focussing on the sheath dynamics and the IEDFs. 
We conclude that the IEDFs are completely
controlled by the sheath dynamics, even in magnetized CCRFDs. Having
argued that, we show that EST can be used as an efficient postprocessing
tool to obtain the IEDFs behind the sheaths in non-magnetized and
magnetized discharges \cite{prenzel2012}.

\section{Discharge Setup}\label{sec:dischargesetup}

Magnetized discharges used for processes such as MERIE employ 
a static and usually inhomogeneous magnetic field to obtain a 
high electron density at very low gas pressures (Pascal or 
sub-Pascal range). The spatial dimensions of such discharges 
are typically a few centimeters, and the magnetic field strength 
can vary up to tens of mT. Under such conditions the electrons 
are magnetized and their effective mobility is reduced; as will 
be discussed, later this leads to a larger voltage drop across 
the plasma bulk and thus to a larger plasma density and 
corresponding ion flux. The ions, in contrast, are \textit{not} 
magnetized (particularly not in the sheath), as their Larmor 
radius $r_L$ is large compared with the typical length scales 
of the system. For instance, the ratio of $r_L$ and the average 
sheath width $\bar{s}$ is $r_L/\bar{s} = \omega_{pi}/\Omega_i\sim 100$. 
Here $\omega_{pi}$ and $\Omega_i$ are the ion plasma frequency and 
the ion cyclotron frequency, respectively.

\begin{figure}[t!]
\centering
\resizebox{0.95\columnwidth}{!}{\includegraphics{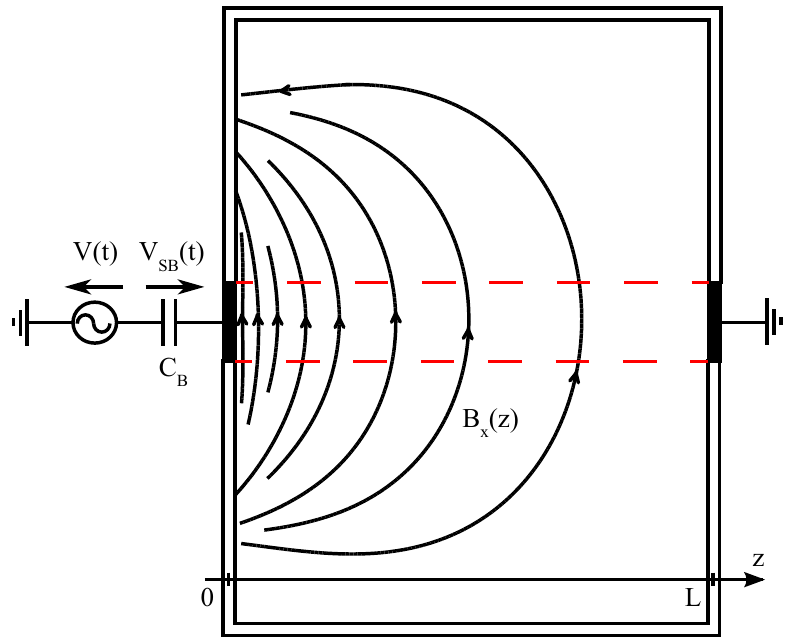}}
\caption{Schematic of the discharge configuration. A static 
magnetic field is assumed at the left electrode which falls 
off along the $z$ axis according to \eqref{eq:fielddecay}. 
There are two symmetric electrodes (solid black) and the plasma 
is confined by a dielectric wall. The dashed 
lines indicate the region of interest for this work.}
\label{fig:geometry}
\end{figure}

In this work we investigate a magnetized Argon discharge at $p=1$ Pa.
The chamber configuration is depicted in figure \ref{fig:geometry}; 
it is a flat, geometrically symmetric cylinder of height $L$ = 5 cm. 
The chamber walls are dielectric, except for the two opposite 
electrodes of area 10$^{-4}\pi$ m$^2$ (which is small compared to 
the diameter of the chamber). The geometric symmetry of the chamber 
is broken by an external magnetic field of ``half-magnetron'' shape 
which is applied near the left electrode (cf. the magnetic field
configuration depicted in figure \ref{fig:geometry}), and by an 
electrically asymmetric discharge voltage $V(t)$ applied to the
left electrode via a blocking capacitor of $C_{B} = 1.5$ nF. 
We assume the following explicit form, where the relative phase 
$\Delta\varphi$ -- the control parameter for the Electrical Asymmetry 
Effect (EAE) -- varies between $-\pi/4$ and $\pi/4$:
\begin{gather}
V(t) = 250\ \mbox{V} \big(\sin(2 \pi \times 13.56\ \mbox{MHz}\ t
+ \Delta\varphi) +  \sin(4 \pi \times 13.56\ \mbox{MHz}\  t)
\big).
\end{gather}
 
This discharge is simulated under the assumption of a one-dimensional
geometry. Clearly, this assumption is rather drastic and neglects the 
inherent inhomogeneity of the magnetic field and many phenomena related 
to it (e.g., particle drifts and magnetic mirror effects). However, 
for the purpose of calculating IEDFs behind the sheaths the assumption 
can be justified, at least for the small corridor (indicated by the 
dashed lines) where the magnetic field is approximately parallel 
to the electrodes.
 
\begin{figure}[t!]
\centering
\resizebox{0.95\columnwidth}{!}{\includegraphics{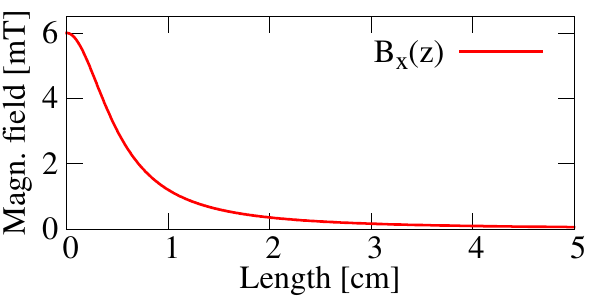}}
\caption{Magnetic field profile inside the discharge.}
\label{fig:magneticfield}
\end{figure}

In a magnetized discharge, current flow along the field lines is not
inhibited. However, in our case, all field lines end at dielectric
sections of the wall; this suppresses the RF current and turns the 
field lines into equipotential lines. In other words, the current can 
only flow into the electrodes, roughly along the dashed corridor. 
In our one-dimensional model we assume that the magnetic field is 
given as follows:
\begin{gather}
\vec{B} = \frac{B_0 \hat{x}}{1+\left( z / l_B \right)^2}
\label{eq:fielddecay},
\end{gather}
where the maximum magnetic field amplitude is $B_0$ = 6 mT and the
decay length is $l_B$ = 5 mm. The profile of $B_x(z)$ within the 
corridor is shown in figure \ref{fig:magneticfield}.

Preliminary results of a self-consistent two-dimensional PIC
simulation of the discussed discharge configuration indicate
that the presented reasoning is essentially valid: a one-dimensional
approximation can capture the essential physics relevant to our topic.

\section{Numerical Analysis}\label{sec:numericalanalysis}

To analyze the magnetized (and non-magnetized) CCRFD described above,
we employ two conceptually different kinetic approaches.
The first is the explicit electrostatic 1d3v PIC code \textit{yapic} 
which has been benchmarked against a number of different PIC 
codes \cite{turner2013}. Within a single PIC cycle the Poisson 
solver, the particle pusher and the Monte-Carlo collision 
module are successively iterated until convergence of the overall 
simulation is achieved. To couple the different modules, a first 
order field interpolation and charge assignment is performed. 
(See e.g., \cite{birdsall1991}). Since the discharge is assumed 
to be magnetized both particle species, electrons and ions, are 
moved according to the full Lorentz force law
$\vec F= \pm q ( \vec E + \vec v\times \vec B )$, although the 
ions are mainly non-magnetized due to their large mass). We 
apply an explicit push scheme based on Boris' approach 
\cite{boris1970}. Collisions of electrons and ions with the 
fixed background gas are performed in the frame of a slightly 
modified null-collisions method reported by Mertmann et al. 
\cite{mertmann2011}, with respect to the classical method
proposed by Skullerud \cite{skullerud1968}. Since Argon is used as
the background gas the collision processes include elastic scattering
of both electrons and ions, ionization and excitation due to electron
collisions, and backward scattering (i.e., charge exchange collisions) 
of ions. Cross-section data are taken from Phelps et al. and the LXCat
Database \cite{phelps1994, lxcat2012}.

The second approach which is used in the end of this paper 
relies on the Ensemble-in-Space time (EST) scheme which resolves 
only the sheath, unlike PIC which treats the entire discharge.
EST is an iterative algorithm based on the solution
of a set of kinetic equations for the ions, Boltzmann's relation
for the electrons, and Poisson's equation for the electric field.
Similar to PIC, EST employs the null-collision method to account
for elastic and charge exchange collisions of ions, but does not
take into account ionization and excitation processes. 
EST is fed by two input parameters: the first is the ion flux 
through the sheath. The second is the sheath voltage which
can have an arbitrary but periodic waveform. Because of its efficient
convergence behavior, EST allows for fast and kinetically self-consistent
simulations of sheaths of CCRFDs. More details concerning the
mathematical description and the validation of the model can be found 
in \cite{shihab2012}.

In order to calculate the sheath voltage as one input parameter 
for EST from the PIC simulation results, we define the time-dependent 
sheath width $s(t)$ using
\begin{gather}
\int_0^{s(t)} n_e dx = \int_{s(t)}^\infty 
\left( n_i - n_e \right) dx.
\end{gather}
This definition is proposed by Brinkmann \cite{brinkmann2007} and
can be substituted by other definitions of the sheath edge.
The sheath voltage is then the potential difference between the
potential at the position $s(t)$ and the electrode potential.
The ion flux as the second input parameter is taken from
the PIC simulation somewhere in the quasineutral zone near the
sheaths. It is clear that the input parameters for EST can
also be obtained from other plasma simulation approaches,
e.g., fluid simulations or global plasma models. It is
important to note that EST does not include
any magnetic field forces.

\section{Results and Discussion}\label{sec:results}

\subsection{Magnetically induced asymmetry of CCRFD}\label{sec:magnetizedpic}

\begin{figure}[t!]
\centering
\resizebox{0.95\columnwidth}{!}{\includegraphics{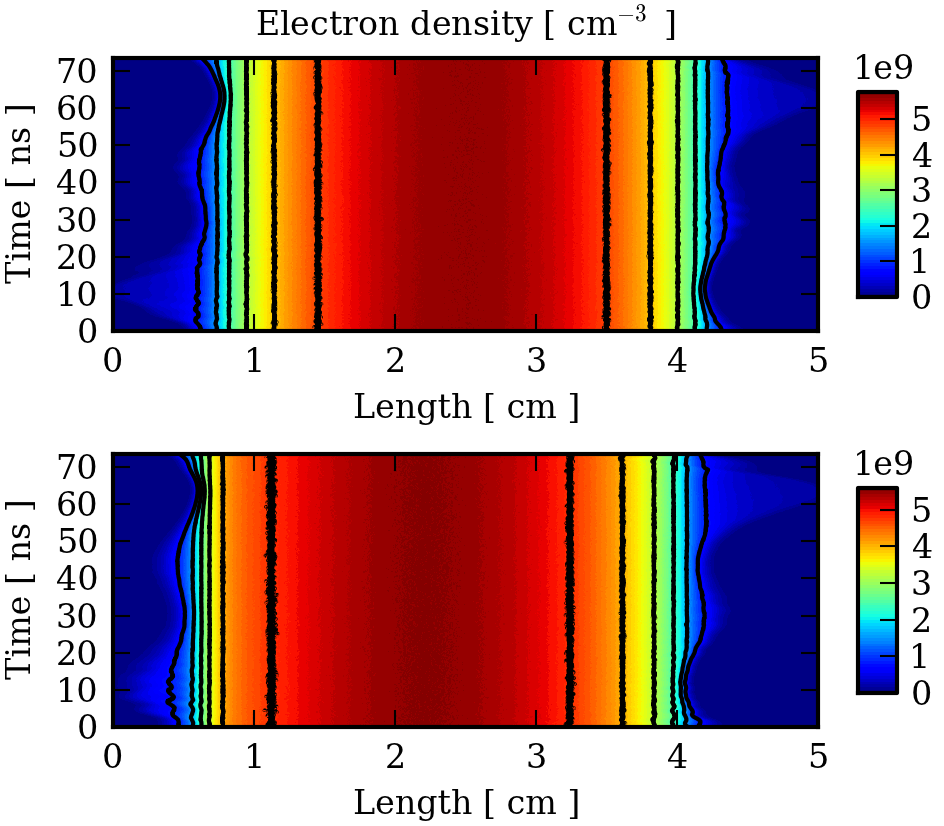}}
\caption{Electron density obtained using the PIC method 
without (top, case I) and with (bottom, case II) an external
magnetic field.}
\label{fig:electrondensity_ab}
\end{figure}

\begin{figure}[t!]
\centering
\resizebox{0.95\columnwidth}{!}{\includegraphics{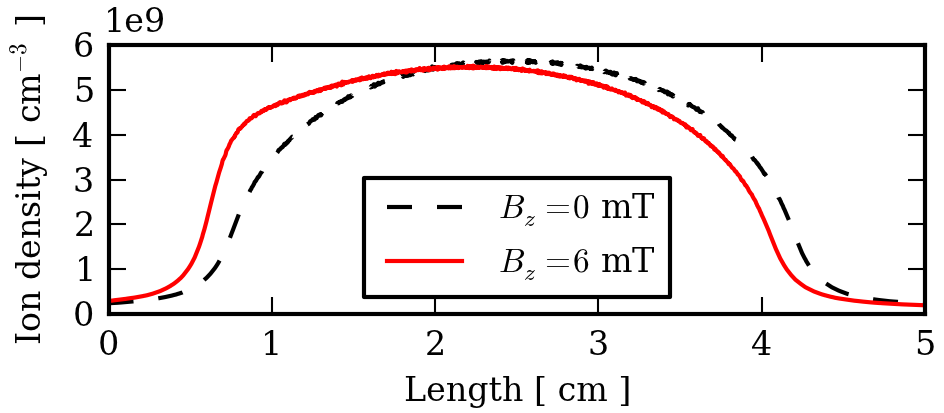}}
\caption{Time-averaged ion density without (dashed, case I)
and with (solid, case II) magnetic field. The maximum is
shifted to the left towards the magnetized plasma sheath.}
\label{fig:iondensity_ab}
\end{figure}

We start our analysis by comparing two different discharge 
scenarios using PIC simulations. Case I is without and case II is 
with an applied static magnetic field at the left electrode.
The simulation parameters are specified in 
section \ref{sec:dischargesetup}. The relative phase
between the two consecutive harmonics of the driving 
voltage is set to $\Delta\varphi=0$. 
Figure \ref{fig:electrondensity_ab} shows the spacetime
dynamics of the electron densities for the two different cases.
For case I we obtain, as expected, a symmetric density profile
and  symmetric sheath dynamics. In contrast, for case II
the discharge shows a significant asymmetry due to 
the applied magnetic field. This asymmetry can also
be observed in figure \ref{fig:iondensity_ab} where the
time-averaged ion density profiles are shown. It is clearly
visible that the maximum is shifted and the slope is steeper
towards the magnetized region in front of the left electrode.

There are two competing mechanisms governing the electrical
conductivity in the magnetized region. At first sight electron 
transport is suppressed by the magnetic field. This leads
to a decrease of the conductivity. On the other hand, due to the 
confinement of the electrons the plasma heating and thus the 
ionization processes become more efficient.
As a result, the plasma density slightly increases locally, 
followed by corresponding enhancement in the conductivity. 
The cumulative effect of these two different mechanisms is that locally
the net conductivity decreases, while the plasma density increases.

Due to the magnetically induced asymmetry, one can also
observe that the sheath in front of the left
driven electrode (which is in the magnetized region)
is much smaller than the sheath
in front of the right grounded electrode. The asymmetry of the 
electron density profile (and the ion flux) is the reason
for the different sheath widths.

\begin{figure}[t!]
\centering
\resizebox{0.95\columnwidth}{!}{\includegraphics{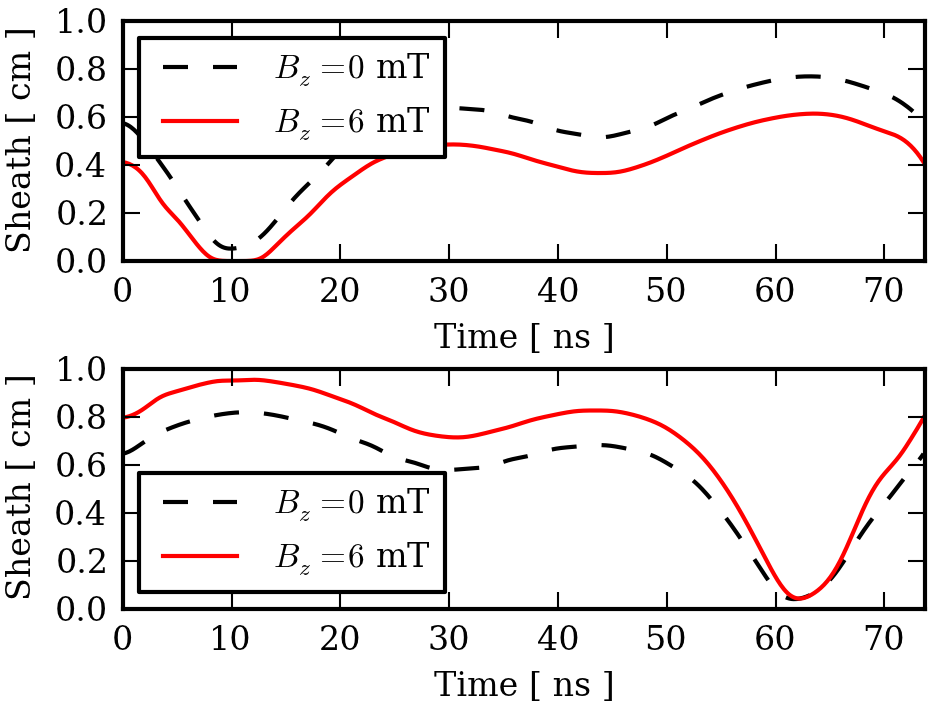}}
\caption{Sheath modulation $s(t)$ (non-magnetized and 
magnetized case, respectively) for the left (top) and 
the right plasma
boundary sheath (bottom).}
\label{fig:sheathwidth_ab}
\end{figure}

\begin{figure}[t!]
\centering
\resizebox{0.95\columnwidth}{!}{\includegraphics{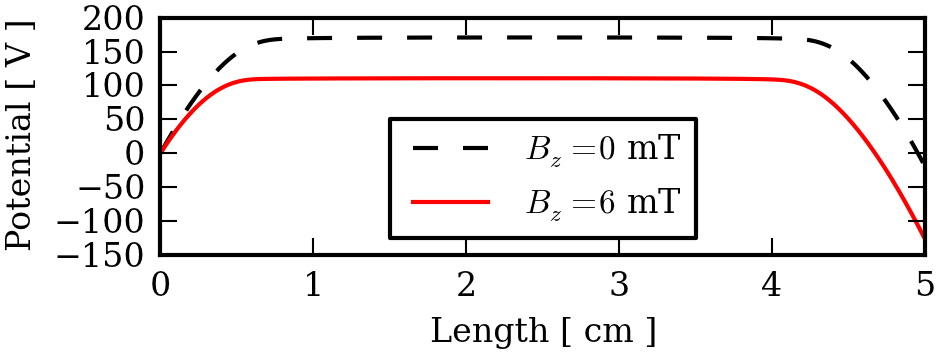}}
\caption{Time-averaged potential along the discharge 
without (dashed, case I) and with (solid, case II) magnetic field.}
\label{fig:potential_ab}
\end{figure}

Figure \ref{fig:sheathwidth_ab} compares the 
temporal behavior of the sheath widths for the
two different cases of both the driven (left) and
grounded (right) sheath. While for case I both sheaths
show very similar dynamics, for case II the left magnetized
sheath is on average significantly smaller than the
right non-magnetized sheath. This magnetically induced asymmetry
results in a substantial DC self bias voltage (of about 130 V).
This can be seen in figure \ref{fig:potential_ab}, where
the time-averaged potential profiles for the two different 
cases are compared. One can also find that the plasma potential
is reduced from about 170 V (for the non-magnetized case) to
about 100 V (for the magnetized case), which is expected.

\begin{figure}[t]
\centering
\resizebox{0.95\columnwidth}{!}{\includegraphics{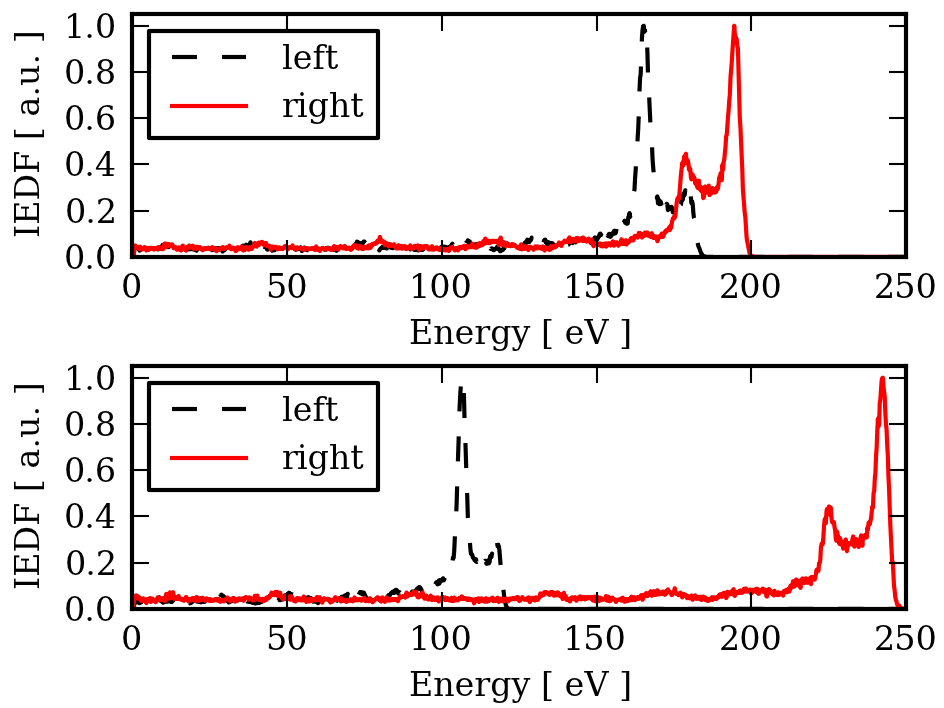}}
\caption{IEDFs for the two cases without (top, case I) and with (bottom, case II) 
an external magnetic field.}
\label{fig:iedf_ab}
\end{figure}

Figure \ref{fig:iedf_ab} shows
energy distribution functions of ions impinging the electrodes
for both cases, without (top, case I) and with (bottom, case II) 
an external magnetic field. With a magnetic field the
discharge is strongly asymmetric. One obtains very high ion energies
at the right grounded electrode while the ion energy is smaller
at the left driven electrode.  This is clearly due to the
DC self bias voltage mentioned before. For the non-magnetized symmetric case
the ion energies at the left and the right electrode are almost equal.
One can observe a small asymmetry in the IEDFs which is due to a
small asymmetry in the sheath potentials. The reason for this is
the difference of the sheath expansion and the resulting beams
of energetic electrons. This effect has been previously reported by 
Schulze et al. \cite{schulze2010}.

\subsection{Electrical Asymmetry Effect in Magnetized CCRFD}\label{sec:magnetizedeae}

The parameter which controls the EAE
and therefore the IEDFs is the relative phase between
the two consecutive harmonics of the driving voltage $\Delta\varphi=0$.
Using a spatially inhomogeneous magnetic field, which
itself produces an asymmetry (as discussed above), one is able
to modify the asymmetry using the EAE. One can argue, that
the EAE can be used for the compensation or amplification of the asymmetry induced
by the inhomogeneous magnetic field. However, in any case the two
effects compete against each other. Here, we briefly discuss
numerical results for case II (the magnetized case) for different
relative phases, i.e., $\Delta\varphi = -\pi/4$, $\Delta\varphi = 0$,
and $\Delta\varphi = +\pi/4$. 

\begin{figure}[t]
\centering
\resizebox{0.95\columnwidth}{!}{\includegraphics{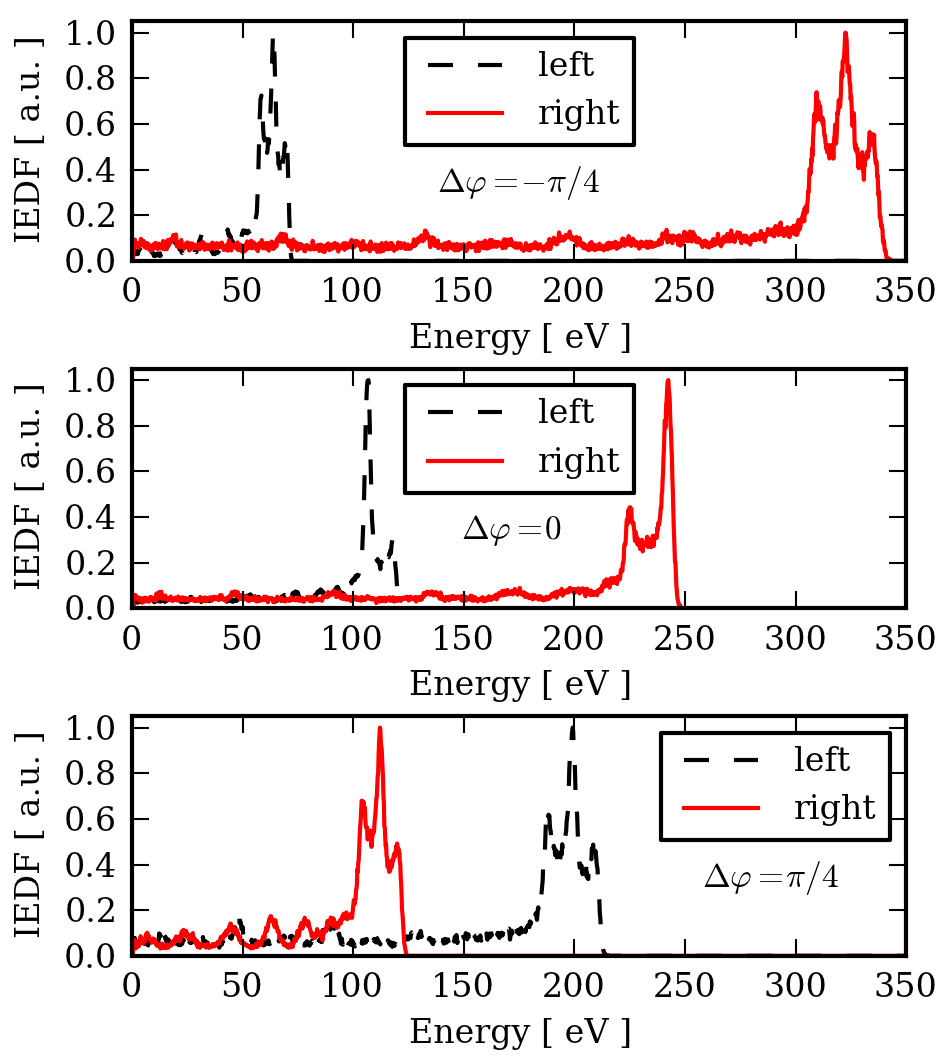}}
\caption{IEDFs for the magnetized case for different 
relative phases of $\Delta \varphi$. The magnetically
induced asymmetry ($\Delta \varphi=0$, middle) can be clearly 
ampflifies ($\Delta \varphi=-\pi/4$, top) and also over-compensated 
($\Delta \varphi=\pi/4$, top) using the EAE.}
\label{fig:iedf_bc}
\end{figure}

Figure \ref{fig:iedf_bc} shows the
IEDFs for this scenario. One can observe that for
$\Delta\varphi = -\pi/4$ the EAE can be used to amplify the
magnetically induced asymmetry (in terms of increasing the
ion bombardment energy at the right grounded electrode and its decreasing
at the same time at the left driven electrode). For a relative phase 
of $\Delta\varphi = \pi/4$ one obtains an over-compensation of
the magnetically induced asymmetry. The role of the left and the
right electrode changes. It is interesting to note that another knob
for controlling the IEDFs could be the change of the absolute value
of the magnetic field in conjunction with the EAE. However, this should
not be discussed in the frame of a simple one-dimensional model, but
should be studied by means of a more complete, at least two-dimensional
model.

\subsection{Influence of the magnetic field on the sheath dynamics}\label{sec:magnetizedest}

In contrast to the electrons, the ions are not
confined by the magnetic field. This is due to their
large mass and a relatively low ion velocity. The main
effect introduced by the external magnetic field is the efficient
electron heating in the region where the magnetic field is high.
When comparing the energy distribution functions of ions impinging
the electrodes for the magnetized and the non-magnetized case, we
can attribute the differences mainly to the different plasma density 
profiles (thus the ion flux towards the electrodes) and to the 
sheath voltages accelerating the ions. The IEDFs are mainly controlled
by the sheath dynamics.

To justify this hypothesis we analyze the sheath dynamics
using EST, which does \textit{not} include any magnetic field effects. 
For this purpose we first compare the spatiotemporal sheath dynamics
obtained from the PIC simulation for the magnetized case (case II) with the results
from EST. As described in section \ref{sec:numericalanalysis} we
extract the ion flux and the sheath voltage from the PIC simulation
and use then as input parameters for EST.

\begin{figure}[t]
\centering
\resizebox{0.95\columnwidth}{!}{\includegraphics{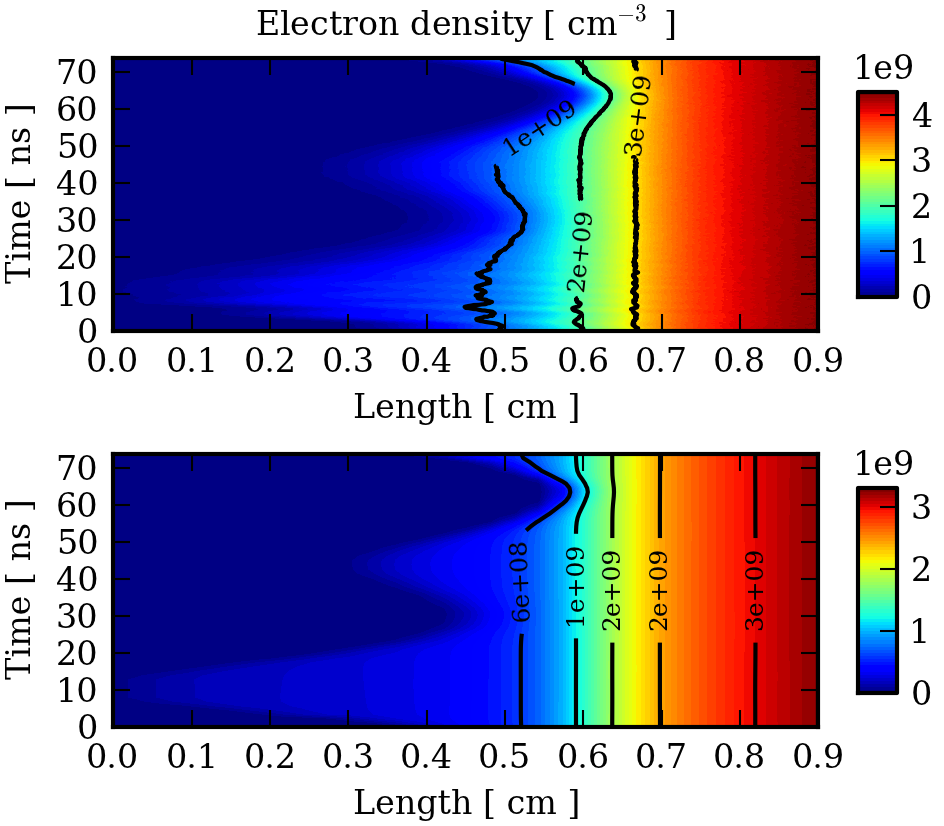}}
\caption{Electron density in the left sheath region obtained 
with the PIC method (top) and the EST model (bottom) for the magnetized
case (case II). Despite the discrepancy in magnitude, there is good 
qualitative agreement regarding the sheath dynamics.}
\label{fig:electrondensity_b_picest}
\end{figure}

Figure \ref{fig:electrondensity_b_picest} shows
the spacetime dynamics of the electron density for both
approaches. The sheath dynamics are in very good qualitative
agreement. Deviations can be observed particularly near
the sheath edge. This is due to the different collision models
used in PIC and EST. While the PIC scheme allows for ionization
collisions of electrons with the background gas, EST does
not include plasma generation. Therefore one can observe a 
higher electron density in the PIC results. A second reason
(which has to be proved) is nonlinear electron resonance heating
due to the excitation of higher harmonics in the plasma current
and thus the plasma series resonance \cite{mussenbrock2006a,
mussenbrock2006b, czarnetzki2006, mussenbrock2007}.
The higher harmonics are clearly visible in the sheath dynamics.
Of course, this phenomenon due to nonlinear sheath bulk interaction
is not included in EST results, since EST resolves the sheath only.

\begin{figure}[t]
\centering
\resizebox{0.95\columnwidth}{!}{\includegraphics{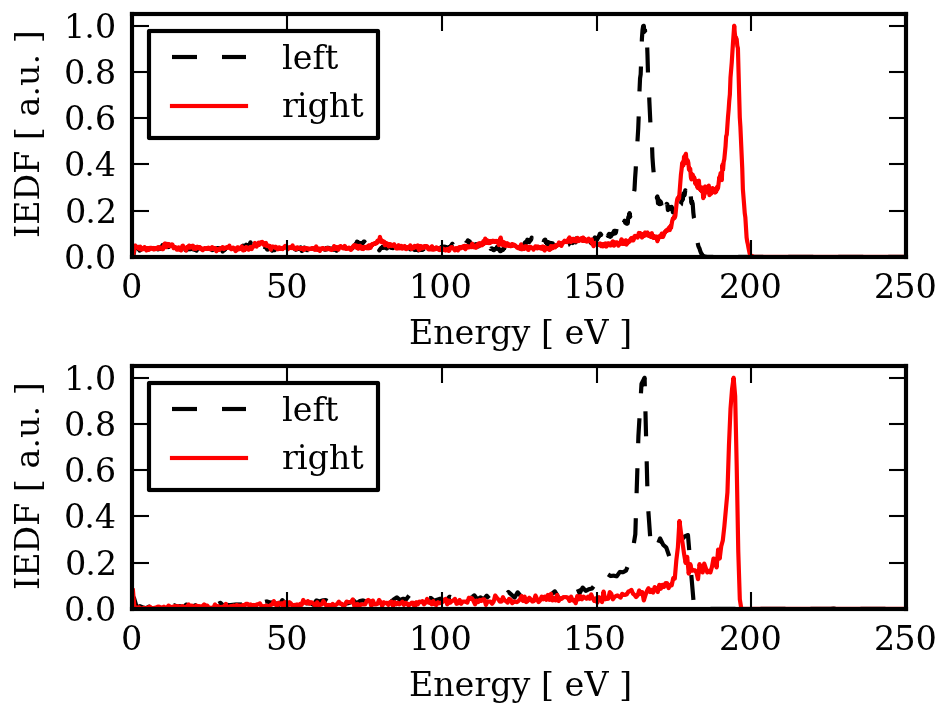}}
\caption{IEDFs obtained for the non-magnetized case (case I)  and  $\Delta\varphi = 0$ using the PIC (top) and the EST model (bottom). There is excellent
agreement between both models in the non-magnetized scenario.}
\label{fig:iedf_a_picest}
\end{figure}

\begin{figure}[t]
\centering
\resizebox{0.95\columnwidth}{!}{\includegraphics{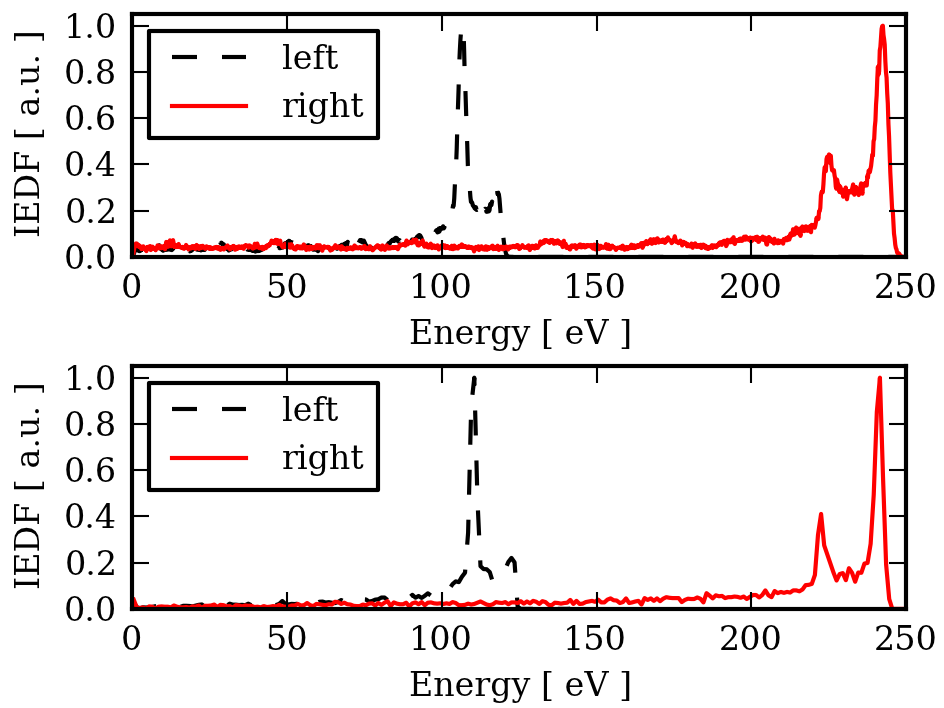}}
\caption{IEDFs for the magnetized case (case II) and 
$\Delta\varphi = 0$. The PIC simulation (top) 
is performed with a constant inhomogeneous magnetic field,
while the EST model (bottom) does not explicitly include the magnetic field, 
but is fed with input data obtained from the magnetic PIC simulations.}
\label{fig:iedf_b_picest}
\end{figure}

Figure \ref{fig:iedf_a_picest} compares
the IEDFs for the non-magnetized case I obtained using both
PIC and EST. It is evident that both models are in
excellent agreement, which is expected. Figure 
\ref{fig:iedf_b_picest} shows the IEDF for the magnetized case II.
The results are again obtained using both PIC and EST. Also
for this case the results are in excellent agreement. Only minor
deviations can be observed when comparing the magnitudes of the
double peak structure.

It is interesting to note that EST yields the same IEDFs
as PIC, regardless of whether the modeled sheath is magnetized or not. 
The reason is, that both the sheath dynamics
and the ion motion in the sheath are mainly driven by the
electric field, and only weakly affected by the magnetic field.
We therefore argue that EST can be used for calculating IEDFs
regardless of the nature of the discharge.

\section{Conclusions}\label{sec:conclusion}

In this work we analyze the effect of a 
magnetic field on the characteristics 
of capacitively coupled radio frequency discharges, 
with a focus on the energy distribution function (IEDF) of 
ions impinging the electrodes and the sheath dynamics. 
Using a well-justified one-dimensional Particle-in-Cell
approach with Monte-Carlo collisions (PIC), we show that 
geometrically symmetric capacitively coupled
radio frequency discharges can be asymmetrized
by applying a spatially inhomogeneous static magnetic
field. This magnetically induced effect is similar to
the electrical asymmetry effect (EAE). We further show that 
the EAE itself can be (over-)compensated and even 
amplified. It this context it could be interesting to
study the effect of the magnitude and the shape of
the external magnetic field on the Electrical Asymmetry
Effect in terms of the symmetry parameter introduced by
Heil et al. \cite{heil2008, czarnetzki2009}. 
The symmetry parameter is defined as the absolute value of 
the ratio of the two sheath voltages. It can be expressed
in terms of an algebraic model. However, it should be discussed in 
detail in the frame of an at least two-dimensional
Particle-in-Cell code which is able to capture the whole 
discharge dynamics including the inherent inhomogeneity
of the magnetic field and the phenomena related to it 
(e.g., particle drifts and magnetic mirror effects).

Finally, we find that the novel Ensemble-in-Space
time scheme (EST) which resolves the sheath only, is
able to provide IEDFs almost equal to the ones obtained
from the fully self-consistent PIC simulation, regardless
of whether the related discharge is magnetized or non-magnetized.
As such EST may be used as an efficient postprocessing tool
to obtain the IEDFs in many scenarios including non-magnetized
and magnetized situations.

\clearpage

\section*{Acknowledgments}
This work is supported by the Deutsche Forschungsgemeinschaft 
DFG in the frame of Collaborative Research Centre TRR 87.

\clearpage

\clearpage

\clearpage

\clearpage

\clearpage

\clearpage

\clearpage

\clearpage

\clearpage

\clearpage

\clearpage

\end{document}